\documentclass[twocolumn,prl]{revtex4-2}

\usepackage{graphicx}
\usepackage{dcolumn}
\usepackage{bm}
\usepackage{multirow}
\usepackage{color}
\usepackage[normalem]{ulem}
\usepackage{amsmath}
\usepackage{pgf}
\usepackage{float}
\usepackage{siunitx}
\usepackage{lipsum}
\usepackage[colorlinks=true]{hyperref}

\usepackage{lineno}
\begin{document}

\title{Probing Nonlinear Interactions of Dipolar Interlayer Excitons in MoSe$_2$/WSe$_2$ Heterobilayers}
\author{Sai Shradha$^{1}$}
\email{sai.shradha@pkm.tu-darmstadt.de}
\author{Luc F. Oswald$^{1}$}
\author{Md Tarik Hossain$^{2}$}
\author{Lukas Krelle$^{1}$}
\author{Nicole Engel$^{1}$}
\author{Axel Printschler$^{2}$}
\author{Julian Führer$^{1}$}
\author{Honey Jayeshkumar Shah$^{2}$}
\author{Daria I. Markina$^{1}$}
\author{Kenji Watanabe$^{3}$}
\author{Takashi Taniguchi$^{4}$}
\author{Andrey Turchanin$^{2}$}
\author{Bernhard Urbaszek$^1$}
\email{bernhard.urbaszek@pkm.tu-darmstadt.de}

\affiliation{\small$^1$Institute for Condensed Matter Physics, TU Darmstadt, Hochschulstraße 6-8, D-64289 Darmstadt, Germany}
\affiliation{\small$^2$Institute of Physical Chemistry, Friedrich Schiller University Jena, Lessingstr. 10, D-07743 Jena, Germany}
\affiliation{\small$^3$Research Center for Electronic and Optical Materials, National Institute for Materials Science, 1-1 Namiki, Tsukuba 305-0044, Japan}
\affiliation{\small$^4$Research Center for Materials Nanoarchitectonics, National Institute for Materials Science,  1-1 Namiki, Tsukuba 305-0044, Japan}

\begin{abstract}
Interlayer excitons in transition-metal dichalcogenide heterobilayers possess intrinsic out-of-plane dipole moments, providing a platform for investigating exciton--exciton interactions at high densities. Here, we use excitation-energy-dependent photoluminescence excitation (PLE) spectroscopy to probe the nonlinear response of dipolar interlayer excitons in chemical vapor deposition-grown MoSe$_2$/WSe$_2$ heterobilayers. By tuning the excitation energy across intralayer exciton resonances at fixed excitation power, we selectively vary the population injected into the interlayer-exciton states. Resonant excitation drives the system into a nonlinear regime, leading to saturation of the interlayer exciton photoluminescence and an apparent broadening of the intralayer $1s$ resonances in the PLE spectra. At the same time, the interlayer exciton emission exhibits a pronounced blueshift, reaching approximately 2.5~meV at 4~K and 1~meV at 75~K. The blueshift increases systematically with the interlayer-exciton population and is consistent with a net repulsive exciton--exciton interaction, with contributions from dipole--dipole repulsion in the density regime investigated. Our results establish PLE as a sensitive approach for accessing the nonlinear, high-density regime of interlayer excitons and probing their interactions in van der Waals heterostructures.
\end{abstract}

\maketitle

\textbf{Introduction.---} 
Heterobilayers of transition metal dichalcogenides (TMDs), such as
MoSe$_2$/WSe$_2$, host properties that
are distinct from those of their constituent monolayers \cite{regan2022emerging,jiang2021interlayer}. The
type-II band alignment in MoSe$_2$/WSe$_2$ spatially separates the
electron and hole into the MoSe$_2$ and WSe$_2$ layers, respectively,
giving rise to spatially indirect interlayer excitons (IXs) with an intrinsic
out-of-plane dipole moment \cite{fang2023,ciarrocchi2019polarization, jauregui2019electrical}. Despite the reduced electron--hole
wavefunction overlap, the Coulomb interaction remains sufficiently
strong to support tightly bound IXs \cite{Rivera2015, kunstmann2018momentum}, while
their spatial separation leads to comparatively long radiative
lifetimes, reaching the nanosecond range \cite{miller2017long,mondal2026enormous}. Together with the
ability to tune their properties through parameters such as twist
angle \cite{choi2021twist,palekar2024anomalous,mahdikhanysarvejahany2021temperature}, stacking configuration \cite{hsu2019tailoring, rivera2018interlayer}, and interlayer spacing \cite{reho2024excitonic}, these
characteristics make dipolar IXs a promising platform
for exploring many-body phenomena, including correlated states and
collective phases \cite{lozovik1976new, wang2019evidence, regan2020mott, li2020dipolar}.  

Accessing such many-body physics requires the generation of sufficiently
dense exciton populations. At elevated densities, interactions between
IXs increasingly influence their energy and relaxation
dynamics, giving rise to a density-dependent spectral response
\cite{Steinhoff2024}. In particular, the permanent dipole moments of
spatially indirect excitons lead to repulsive dipole--dipole
interactions, which are expected to increase the IX
transition energy with increasing exciton density. At the same time,
high exciton densities can introduce nonlinear relaxation and
recombination pathways, making the optical response dependent not only
on the exciton population but also on the processes governing their
formation and decay. Probing the high-density regime therefore
requires both efficient generation of IXs and a
spectroscopic approach that is sensitive to their excitation and
relaxation pathways.

\begin{figure*}
    \centering
    \includegraphics[width=1\linewidth]{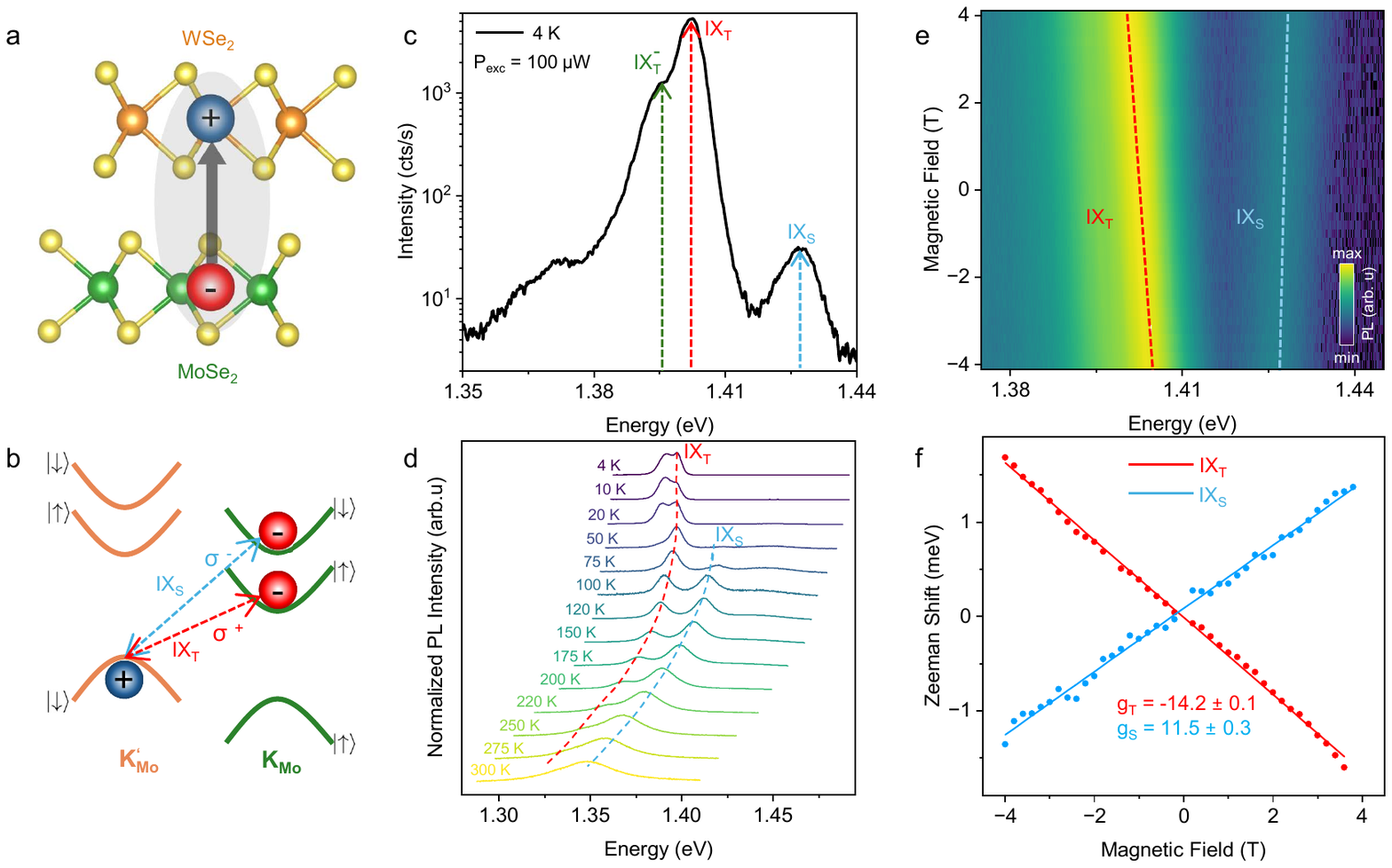}
    \caption{\textbf{a} Schematic of MoSe$_2$/WSe$_2$ heterobilayer. \textbf{b} Band alignment and spin configuration of the type-II MoSe$_\text{2}$/WSe$_\text{2}$ heterostructure, illustrating interlayer charge transfer and the formation of spin-singlet (IX$_\text{S}$) and spin-triplet (IX$_\text{T}$) IXs. \textbf{c} PL spectrum from the heterobilayer resolving multiple interlayer exciton resonances such as the spin-triplet trion $\mathrm{IX}^{-}_\mathrm{T}$, the spin-triplet $\mathrm{IX}_\mathrm{T}$ and the spin-singlet $\mathrm{IX}_\mathrm{S}$ states at T = 4 K. \textbf{d} Temperature-dependent PL spectra ranging from 4 to 300 K showing the energy shift of the PL peaks following a Varshni function depicted with dashed curves. \textbf{e} Magnetic field dependent  shift in the emission energy of $\mathrm{IX}_\mathrm{T}$ and $\mathrm{IX}_\mathrm{S}$. \textbf{f} Zeeman shifts and the corresponding g-factors of the $\mathrm{IX}_\mathrm{T}$ and $\mathrm{IX}_\mathrm{S}$ establish the H$_h^h$ registry of the heterobilayer.  }
    \label{fig:Figure1}
\end{figure*}

Most MoSe$_2$/WSe$_2$ heterobilayers have traditionally been assembled
by mechanically stacking exfoliated monolayers. Recent advances in
chemical vapor deposition (CVD), however, have enabled the direct
growth of TMD heterostructures \cite{paradisanos2020controlling,Li2023}.
A key advantage of bottom-up growth approaches, such as CVD, 
is the formation of atomically clean and well-defined interfaces in contrast to mechanically 
stacked heterostructures. Such high-quality interfaces provide access to physical phenomena 
that are challenging to realize or investigate in mechanically assembled samples.
In particular, liquid-precursor-grown MoSe$_2$/WSe$_2$ heterobilayers
have been shown to exhibit efficient interlayer charge transfer and
robust IX emission from cryogenic temperatures to room
temperature \cite{hossain2026cvd}. The efficient formation of interlayer
excitons in this material platform provides an opportunity to access
their high-density regime while retaining strong optical signatures.

Here, we use PLE spectroscopy to
investigate the nonlinear response and density-dependent interactions
of IXs in CVD-grown MoSe$_2$/WSe$_2$ heterobilayers \cite{hossain2026cvd}.
By tuning the excitation energy across intralayer excitonic resonances,
we vary the population of excitons injected into the
IX state while keeping the laser excitation power constant. We find
that resonant excitation drives the system into a nonlinear regime in
which the IX photoluminescence (PL) response saturates, giving rise to an
apparent broadening of the 1s resonances in the PLE spectra. At the
same time, the high IX population produces a pronounced
blueshift of the IX emission. The systematic increase
of this blueshift with increasing IX population is
consistent with a net repulsive exciton--exciton interaction and
serves as evidence for dipolar interactions between interlayer
excitons.\\
\indent \textbf{Experimental Results and Discussion.---}

Figure~\ref{fig:Figure1} summarizes the structure and optical
properties of the heterobilayers investigated here. As illustrated in
Figure~\ref{fig:Figure1}a, the type-II band alignment spatially
separates the electron and hole between the MoSe$_2$ and WSe$_2$
layers, forming a spatially indirect exciton with an intrinsic
out-of-plane dipole moment \cite{Rivera2015}. The spin-dependent band structure (Figure~\ref{fig:Figure1}b)
gives rise to two dominant IX species: the
spin-allowed interlayer singlet, IX$_\mathrm{S}$, and the
spin-forbidden interlayer triplet, IX$_\mathrm{T}$ \cite{wietek2024,borel2026photoexcitation,Sahoo2026}. For the
$H_h^h$ stacking configuration considered here, IX$_\mathrm{T}$ is
the lower-energy state and therefore represents the dominant
IX population following relaxation \cite{wang2020}.

This behavior is observed directly in the PL spectrum shown in
Figure~\ref{fig:Figure1}c. The emission is dominated by IX$_\mathrm{T}$,
with a weaker IX$_\mathrm{S}$ contribution at higher energy. The contribution of the higher lying IX$_\mathrm{S}$ state 
is however visible at T = 4~K only for sufficiently high excitation powers \cite{fang2023}. A charged
triplet, IX$_\mathrm{T}^{-}$, is also observed approximately 7~meV
below IX$_\mathrm{T}$. Notably, unlike many mechanically assembled
heterobilayers in which residual intralayer exciton emission can
remain visible, we do not detect measurable intralayer exciton
emission from the heterobilayer region within our signal-to-noise
level. The absence of detectable intralayer emission indicates highly
efficient interlayer charge transfer and, consequently, efficient
formation of IXs due to the atomically clean interface \cite{hossain2026cvd}.

The robust formation of IXs is further demonstrated by
the temperature-dependent PL shown in Figure~\ref{fig:Figure1}d.
At low temperatures, the emission is dominated by IX$_\mathrm{T}$,
while the relative spectral weight of IX$_\mathrm{S}$ increases with
temperature and becomes dominant at room temperature. Both excitonic
transitions exhibit a temperature-dependent redshift consistent with
the Varshni relation \cite{varshni1967}; the corresponding fitting
parameters are given in Table~I of the Supplementary Information.
Importantly, IX emission remains clearly observable up
to 300~K. This robust emission over a broad temperature range,
together with the absence of detectable intralayer emission, confirms
the efficient interlayer charge transfer and high optical quality of
the heterobilayer interface required for the experiments below.

The stacking configuration is an important additional parameter because
it determines the electronic structure and optical selection rules of
the heterobilayer \cite{Liu2014,seyler2019signatures}. Since both
$2H$- and $3R$-type stacking configurations can occur in CVD-grown
heterostructures \cite{Li2023,paradisanos2020controlling}, we determine the stacking order optically using
magneto-optic PL. An out-of-plane magnetic field produces
valley-dependent Zeeman shifts of the IX transitions.
Figure~\ref{fig:Figure1}e shows the PL as a function of magnetic field,
where IX$_\mathrm{T}$ and IX$_\mathrm{S}$ exhibit shifts in opposite
directions. The corresponding Zeeman g-factors are extracted
according to
\begin{equation}
    \Delta_Z = E_{\sigma^+}-E_{\sigma^-}
             = g\mu_B B_z,
\end{equation}
where $g$ is the effective exciton $g$-factor. As shown in
Figure~\ref{fig:Figure1}f, we obtain
$g_\mathrm{T}= -14.2\pm0.1$ and
$g_\mathrm{S}=11.5\pm0.3$. The magnitude and sign of these
$g$-factors are consistent with the $H_h^h$ stacking configuration
reported for MoSe$_2$/WSe$_2$ heterobilayers
\cite{wietek2024}, thereby establishing the stacking configuration of
the sample used in the subsequent interaction measurements.\\
\begin{figure*}
    \centering
    \includegraphics[width=1\linewidth]{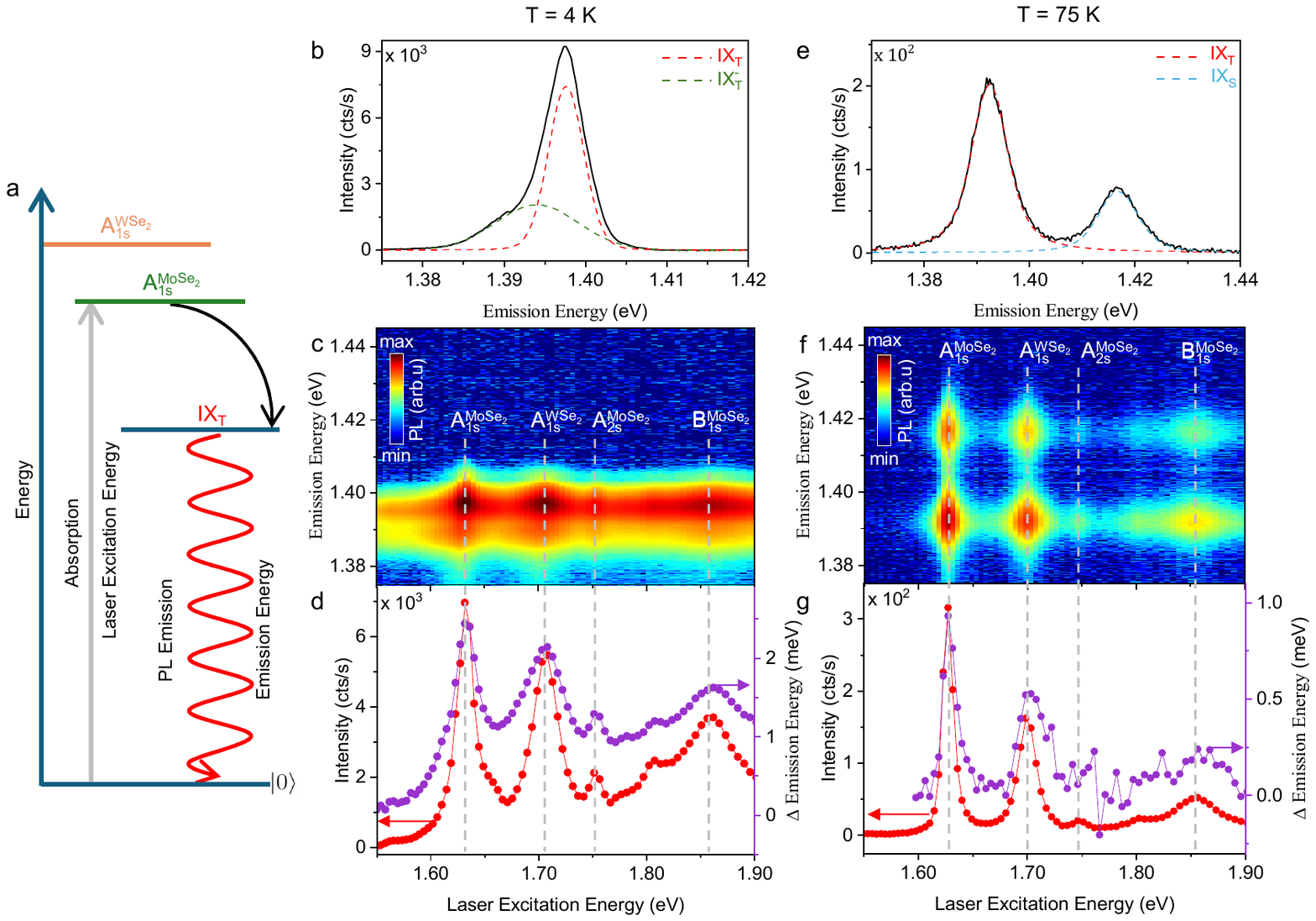}
    \caption{\textbf{a} Schematic of absorption, charge transfer and PL emission under excitation resonant with the intralayer excitonic states. \textbf{b} PL emission under an excitation of 1.63~eV. The green and red dashed curves show the fits corresponding to $\mathrm{IX}^{-}_\mathrm{T}$ and $\mathrm{IX}_\mathrm{T}$ respectively.  \textbf{c} Heatmap of PL intensity while tuning the excitation energy. \textbf{d} Peak intensity (in red, left y-axis) and shift ($\Delta$) of the emission energy (in purple, right y-axis) of $\mathrm{IX}_\mathrm{T}$ as a function of the excitation energy. The grey dashed lines indicate the position of the resonances which are assigned to various intralayer states. \textbf{b-d} are all measured at T = 4~K with a constant excitation power of 20~µW. \textbf{e-g} The same as \textbf{b-d} measured at T = 75~K and also with a laser excitation power of 20~µW. The red and blue dashed lines are the fits respectively assigned to $\mathrm{IX}_\mathrm{T}$ and $\mathrm{IX}_\mathrm{S}$.}
    \label{fig:Figure2}
\end{figure*}
\indent Having established efficient IX formation in the CVD-grown heterobilayer,
we next investigate how their population responds to different excitation pathways using PLE. 
IX states in such heterobilayers, do not carry significant oscillator strengths 
relative to that of intralayer excitons \cite{wietek2024,barre2022optical}. 
This stems mainly from the reduced overlap of the wavefunctions of the spatially separated electrons and holes. 
The formation of IXs can therefore be optimised by efficient absorption into the 
intralayer states that feed it through charge transfer processes. 
This formation of IXs can be probed by PLE spectroscopy where the laser 
excitations energy is tuned through various intralayer resonances while detecting the interlayer emission, see methods.
Figure~\ref{fig:Figure2}a shows the schematic of this process where absorption into intralayer states is 
followed by charge transfer leading to the formation of IX and finally relaxation back to the ground state. 
A typical spectrum under resonant excitation (at the A$_{1s}^{\mathrm{MoSe_2}}$ energy)  at T = 4~K is shown in Figure~\ref{fig:Figure2}b. 
Figure~\ref{fig:Figure2}c presents a heatmap of the PLE response
measured at an excitation power of 20~$\mu$W as a function of the laser excitation energy. Distinct
features are observed at the A$_{1s}^{\mathrm{MoSe_2}}$,
A$_{1s}^{\mathrm{WSe_2}}$, A$_{2s}^{\mathrm{MoSe_2}}$, and
B$_{1s}^{\mathrm{MoSe_2}}$ transitions
\cite{wang2015exciton,stier2018magnetooptics}. Interestingly, the 1s resonances appear
unusually broad in the PLE response. For example, the
A$_{1s}^{\mathrm{MoSe_2}}$ resonance has a FWHM of approximately
19~meV. This is substantially broader than the 6~meV linewidth observed for
hBN-encapsulated monolayer TMDs in absorption (see Figure~S2 of the Supplementary Information) (\cite{wierzbowski2017direct,Cadiz2017}).

It is important to note, however, that PLE does not exclusively probe 
absorption. The detected interlayer PL results from a sequence of 
processes involving absorption into an intralayer
exciton, interlayer charge transfer, relaxation, and radiative
recombination \cite{shree2018observation,klingshirn2007semiconductor}. 
Consequently, the PLE intensity reflects not only the
efficiency of optical excitation, but also the efficiency with which
the resulting exciton population is transferred and converted into
radiative IXs. In particular, under resonant
excitation the strong absorption into the intralayer 1s states can
generate a large population of IXs. At sufficiently
high populations, exciton--exciton interactions can modify both the
energy and the PL yield of the IXs
\cite{Steinhoff2024,perea2022exciton}. Thus, the apparent width of a
PLE resonance need not directly correspond to the intrinsic linewidth
of the underlying intralayer absorption resonance.

To distinguish intrinsic spectral broadening from nonlinear interaction effects in
the excitation-to-emission pathway, we compare the PLE response at
4~K and 75~K. Figures~\ref{fig:Figure2}e-g show the corresponding
measurements at 75~K. Remarkably, the resonances corresponding to 1s states become substantially
narrower at the elevated temperature. The FWHM of the
A$_{1s}^{\mathrm{MoSe_2}}$ resonance decreases from 19~meV at 4~K to
10~meV at 75~K, while that of the A$_{1s}^{\mathrm{WSe_2}}$ resonance
decreases from 30~meV to 18~meV. This corresponds to reductions of
approximately 47\% and 40\%, respectively. Such a trend is
counterintuitive if the observed linewidths were determined primarily
by conventional phonon broadening, which is expected to increase with
temperature \cite{Cadiz2017}. The temperature dependence therefore
suggests that the broadening observed in the low-temperature PLE
spectra is predominantly related to nonlinear population dynamics
rather than to an intrinsic broadening of the 1s absorption resonances.

A second indication of exciton-exciton interactions is provided by
the relative intensities of the A$_{1s}^{\mathrm{MoSe_2}}$ and
A$_{2s}^{\mathrm{MoSe_2}}$ resonances. At 4~K, the A$_{2s}$ feature
reaches more than one third of the A$_{1s}$ intensity, whereas at
75~K the relative strength is substantially reduced and approaches
the expected hierarchy of the excitonic Rydberg series
\cite{chernikov2014exciton,Wang2018,robert2018optical}. This again
suggests that the strong resonant excitation of the 1s state is
driving the IX population into a nonlinear regime.

The origin of this nonlinear response becomes apparent from the
excitation-energy-dependent emission energy. At both temperatures,
the IX$_{\mathrm{T}}$ emission exhibits a pronounced energy shift when
the excitation energy is tuned to an intralayer exciton resonance. 
The purple plots in Figures~\ref{fig:Figure2}d and g show the emission energy shifts relative to that under an off-resonant excitation 
(in this case with a laser excitation energy of 1.53~eV at 4~K and 1.60~eV at 75~K).
The largest shift occurs at the A$_{1s}^{\mathrm{MoSe_2}}$ resonance, reaching approximately 2.5~meV
at 4~K and 1~meV at 75~K. The coincidence of the strongest PLE
response and the largest blueshift at the same excitation resonance
indicates that the resonant excitation generates the highest
IX population. The resulting energy shift therefore
provides a direct spectroscopic fingerprint of the high-density regime.

\begin{figure*}
    \centering
    \includegraphics[width=0.8\linewidth]{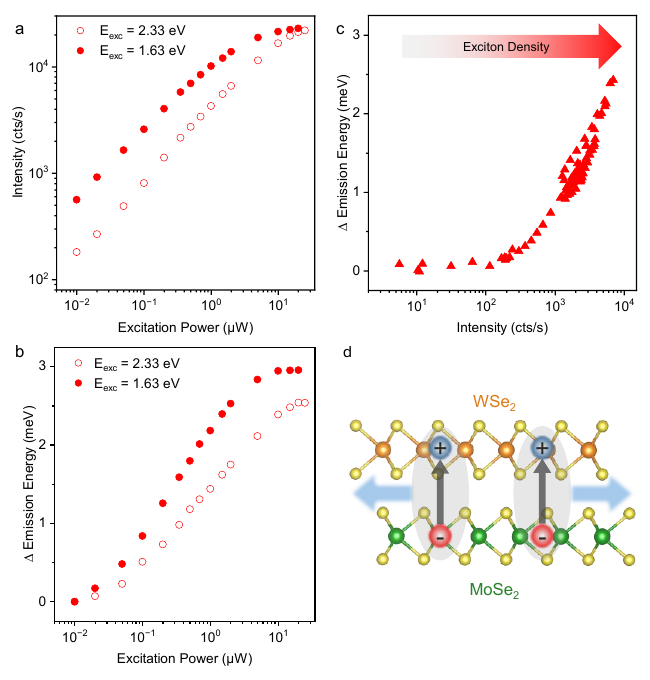}
    \caption{Power dependent \textbf{a} intensity and \textbf{b} emission energy shift, of $\mathrm{IX}_\mathrm{T}$ under resonant (1.63~eV, solid circles) and off-resonant (2.33~eV, circles) excitation. \textbf{c} The shift of the emission energy with corresponding change in the intensity of the of $\mathrm{IX}_\mathrm{T}$ emission obtained from PLE at T = 4~K with a power of 20~µW. \textbf{d} Schematic of dipolar repulsion that dominates the interaction between dipolar IXs at high exciton densities. }
    \label{fig:Figure3}
\end{figure*}

To establish the connection between resonant excitation and exciton
population more directly, we perform power-dependent PL measurements
under resonant A$_{1s}^{\mathrm{MoSe_2}}$ excitation at 1.63~eV and off-resonant excitation at
2.33~eV, as shown in Figure~\ref{fig:Figure3}a and b. Under resonant
excitation, the IX$_{\mathrm{T}}$ PL enters a saturation regime already
at excitation powers of approximately 10~$\mu$W, whereas the
off-resonant PL remains below saturation over the same power range.
This demonstrates that the resonant excitation efficiently drives the
system into a high-exciton-population regime.
The initially injected exciton density can be estimated as
\begin{equation}
    n_e = \frac{P_{\mathrm{exc}}\alpha(E_{\mathrm{exc}})}
    {f_{\mathrm{rep}}\pi r^2 E_{\mathrm{exc}}},
    \label{eq:exciton density}
\end{equation}
where $P_{\mathrm{exc}}$ is the laser excitation power,
$\alpha(E_{\mathrm{exc}})$ is the excitation-energy-dependent
absorption, $f_{\mathrm{rep}}$ is the laser repetition
rate, $r$ is the excitation-spot radius, and $E_{\mathrm{exc}}$ is the
laser excitation energy. While conventional power-dependent measurements
vary the exciton density by changing $P_{\mathrm{exc}}$, PLE provides
a complementary approach: at fixed excitation power, tuning
$E_{\mathrm{exc}}$ changes $\alpha(E_{\mathrm{exc}})$ considerably and therefore
the population of initially injected excitons.

Equation~\ref{eq:exciton density} provides an estimate of the initially
injected exciton density rather than the absolute density of
IX$_{\mathrm{T}}$, since the latter additionally depends on the
efficiency of interlayer charge transfer and the subsequent relaxation
dynamics. We therefore use the IX$_{\mathrm{T}}$ PL intensity as an
experimental proxy, i.e. an uncalibrated dependence, for its population. Figure~\ref{fig:Figure3}c shows
the corresponding IX$_{\mathrm{T}}$ emission-energy shift
($\Delta$ Emission Energy), as a function of the emission intensity.
These are obtained from the PLE measurements shown in Figure~\ref{fig:Figure2}d.
A monotonic increase of the blueshift with increasing emission
intensity is observed over the population range investigated. This
behavior is consistent with a net repulsive interaction between
dipolar IXs. Because the electron and hole are
spatially separated, each IX possesses an out-of-plane
electric dipole moment, leading to dipole--dipole repulsion between
neighboring excitons. As the IX population increases, the mean
inter-exciton separation decreases and the repulsive interaction
becomes stronger, resulting in the observed blueshift. 
As discussed in \cite{Steinhoff2024} such net-blueshift interactions may stem mainly from three interaction types, namely, dipole--dipole repulsion, phase-space-fillinng and bosonic exchange interaction. 
In the density regime investigated here, these contributions appears to
outweigh interaction mechanisms expected to produce a redshift, such
as fermionic exchange interaction and screened bosonic exchange.

The same nonlinear population dynamics also explain the apparent
broadening of the 1s PLE resonances at 4~K. At resonance, the strong
intralayer absorption generates a sufficiently large IX
population that the subsequent IX PL response no longer follows a
linear relationship with the absorbed excitation power. At the
20~$\mu$W excitation power used for the PLE measurements, the resonant
population is already sufficiently high for exciton--exciton
Auger-type scattering to contribute to the relaxation dynamics. The
resonant PLE intensity is consequently suppressed relative to the
linear-response expectation. Because this suppression is strongest
near the center of the 1s absorption resonances, the resulting PLE
features appear artificially broadened. Thus, the broad 1s features
observed in the 4~K PLE spectra should not be interpreted as evidence
for intrinsically broad 1s absorption resonances, but rather as a
signature of the nonlinear high-density IX regime. An additional confirmation of this 
was made via PLE measurents in the linear regime of exciton density. 
This can be see the PLE spectrum with a 100~nW excitation power show in Figure~S1 of the Supplementary Information.\\
\indent \textit{In summary}, we have used excitation-energy-dependent photoluminescence 
excitation spectroscopy to access the nonlinear, high-density regime of interlayer excitons in
chemical vapor deposition-grown MoSe$_2$/WSe$_2$ heterobilayers. By tuning the excitation energy 
at fixed excitation power, we selectively control the population injected into the interlayer-exciton states. 
Resonant excitation generates sufficiently high populations to drive the interlayer-exciton 
photoluminescence into a nonlinear regime, resulting in an apparent broadening of the intralayer $1s$ 
resonances in the PLE spectra. The temperature dependence and power-dependent measurements 
show that this broadening originates predominantly from nonlinear population dynamics 
rather than from intrinsically broad intralayer absorption resonances.

More importantly, increasing the interlayer-exciton population produces a systematic blueshift of 
the interlayer-exciton emission, reaching approximately 2.5~meV at 4~K and 1~meV at 75~K. 
The monotonic dependence of the blueshift on the interlayer-exciton population demonstrates 
a net repulsive exciton--exciton interaction in the density regime investigated.
These results establish excitation-energy-dependent PLE as a sensitive spectroscopic approach for selectively accessing dense interlayer-exciton populations and probing their nonlinear interactions. This provides a route toward investigating many-body phenomena associated with dipolar excitons in van der Waals heterostructures.

\textbf{METHODS}\\
\indent\textbf{Sample Fabrication}\\
The heterobilayer MoSe$_\text{2}$/WSe$_\text{2}$ was grown using a liquid precursor 
based chemical vapor deposition method \cite{hossain2026cvd} on a SiO$_2$/Si substrate. 
The stack was then encapsulted with hBN \cite{Cadiz2017,Shradha2026} using a 
polymer-based hot pick-up technique \cite{Pizzocchero2016}.\\
\indent\textbf{PL and PLE Spectroscopy}\\
A home-built spectroscopy setup \cite{shree2021guide} was used to perform the PL and PLE measurements. 
The sample was placed inside a closed-cycle cryostat (Attocube systems, AttoDry 800) 
on a stack of low-temperature piezo-positioners (Attocube systems, ANPx101 and ANPz102). 
to position the sample with respect to the objective (CryoGlass Optics, 50x, NA = 0.7). 
The objective in the backscattering geometry
focused the laser beam on the sample surface at normal incidence. The spot size diameter was of 
the order of the wavelength used. This was measured experimentally for different wavelengths by 
scanning the reflection signal over a metal stripe. For the PLE measurements a unpolarized white 
light fiber laser (Fianium FIU-15, NKT Photonics) followed by a tunable high-contrast filter (LLTF Contrast VIS HP8, NKT Photonics) providing the spectral line width of $\sim$1~nm were used. The emission spectra were acquired via 
free-space path by a Czerny-Turner spectrometer (Teledyne Princeton Instruments IsoPlane300) 
equipped with a Peltier cooled CCD (Teledyne Princeton Instruments, Blaze 400HRX). In case of the PLE measurements a 850~nm shortpass filter (FESH05850) and a 850~nm longpass filter (FELH0850) were used in the excitation and detection paths respectively.\\
\indent\textbf{Magneto-optic Spectroscopy}\\
The magneto-optic Zeeman shift was measured using a home-built confocal microscope setup 
with the sample placed inside a closed-cycle cryostat (Attocube systems, AttoDry 1000XL) with a vector magnet. 
The emission was filtered with a 650~nm longpass filter (FELH0650) following which it was dispersed through a Czerny-Turner spectrograph (Teledyne Princeton Instruments, SpectraPro HRS-500)
and detected by a CCD (Teledyne Princeton Instruments, Pylon BRexcelon 100).  \\
The signal to noise ratio for all spectral measurements obtained at T = 4~K is in the order of roughly $10^3$. This allows us to extract 
shift considerably smaller than the linewidth of the emission \cite{belhadj2009controlling}. 
%

\textbf{Data availability}\\
The data that support the findings of this study are available from the corresponding authors upon request.\\

 \textbf{Acknowledgements}\\
 
We thank Alexander W. Holleitner, Mauro Brotons-Gisbert and Alexander Högele for insightful discussions on heterobilayer systems. Additionally, we thank Wenze Lan for feedback on the manuscript. We acknowledge that the atomic lattice structures depicted in the manuscript were made using VESTA software \cite{momma2011vesta}.

\textbf{Funding Declaration}\\
A.T, M.T.H, A.P, and H.J.S acknowledge financial support of the DFG through projects TU 149/16-1  (464283495); TU149/21-1 (535253440), 
CRC NOA 1375, B02 (398816777) and European Union Graphene Flagship project 2D Materials of Future 2DSPIN-TECH (101135853). 
K.W and T.T acknowledge support from the JSPS KAKENHI (Grant Numbers 21H05233 and 23H02052) , the CREST (JPMJCR24A5), JST and World
Premier International Research Center Initiative (WPI), MEXT, Japan. 
B.U, A.T and S.Sh acknowledge financial support by the Deutsche Forschungsgemeinschaft  (DFG) via SPP 2244. 
B.U and D.I.M acknowledge funding from LOEWE Exploration project Exziton-basierte technologische Anwendungen.

\textbf{Author contributions} 
M.T.H,  A.P, H.J.S and A.T grew the heterostructures and spectroscopically and microscopically pre-characterized the as-grown samples. K.W and T.T grew the bulk hBN crystals. S.Sh, J.F and N.E fabricated the encapsulated samples for optical spectroscopy. S.Sh, L.F.O,  and L.K performed optical spectroscopy measurements. S.Sh and D.I.M.,installed the cryostat system. S.Sh and L.F.O analyzed the optical spectra. All authors discussed the results. B.U. suggested the experiments and supervised the project. S.Sh and B.U. wrote the manuscript with input from all the authors. \\

\textbf{Competing interests}: The authors declare no competing interests.\\


\clearpage
\onecolumngrid
\section*{Supplementary Information}
\renewcommand{\thefigure}{S\arabic{figure}}
\setcounter{figure}{0}

\vspace{2cm}
\begin{table*}[h]
\centering
\label{tab:varshni_parameters}
\begin{tabular}{|c|c|c|c|}
\hline
Transition & $E_0$ (eV) & $\alpha$ ($10^{-4}$ eV/K) & $\beta$ (K) \\
\hline
$\mathrm{IX_T}$      & 1.400 & 6.1 & 536 \\
$\mathrm{IX_S}$      & 1.425 & 7.0 & 550 \\
MoSe$_2$ (exfoliated) & 1.640 & 8.3 & 597 \\
WSe$_2$ (exfoliated) & 1.720 & 6.0 & 327 \\
\hline
\end{tabular}
\caption{Varshni fit parameters for $\mathrm{IX_T}$ and $\mathrm{IX_S}$ obtained from temperature-dependent PL measurements. Values for monolayer MoSe$_2$ and WSe$_2$ are included for comparison.}
\end{table*}
\vspace{2cm}
\begin{table*}[h]
\centering
\label{tab:PLE_widths}
\begin{tabular}{|c|c|c|c|}
\hline
Temperature & Laser Excitation Power (µW) & A$_{1s}^{\mathrm{MoSe_2}}$ FWHM (meV) & A$_{1s}^{\mathrm{WSe_2}}$ FWHM (meV) \\
\hline
\multirow{2}{*}{\centering 4~K} & 20     & 19  & 30  \\
& 0.1      & 11 & 20  \\
\hline
75~K & 20     & 9  & 18  \\
\hline
\end{tabular}
\caption{Comparison at PLE resonance full width at half maximum (FWHM) for the IX$_{\mathrm{T}}$ emission at different temperatures and excitation powers.}
\end{table*}

\newpage
\begin{figure*}
\includegraphics[width=0.65\linewidth]{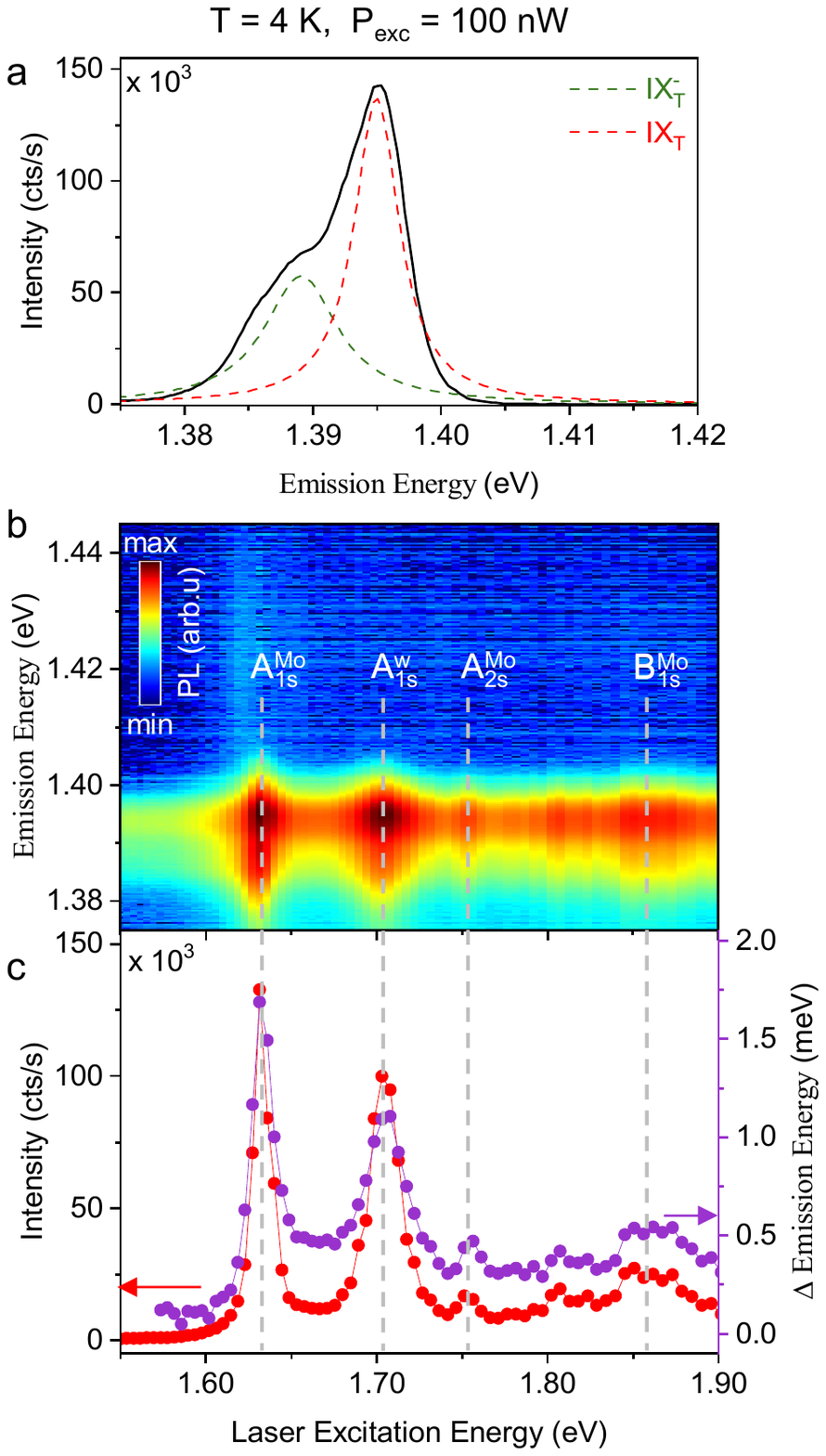}
\caption{\textbf{a} Emission from the heterobilayer region at with laser excitatio energy of 1.63~eV. \textbf{b} PLE heatmap. \textbf{c} Peak intensity (red plot, left y-axis) and emission energy shift(purple plot, right y-axis) of IX$_{\mathrm{T}}$ as a function of the laser excitation energy. }
\label{fig:FigureS1} 
\end{figure*}

\newpage
\begin{figure*}
\includegraphics[width=0.8\linewidth]{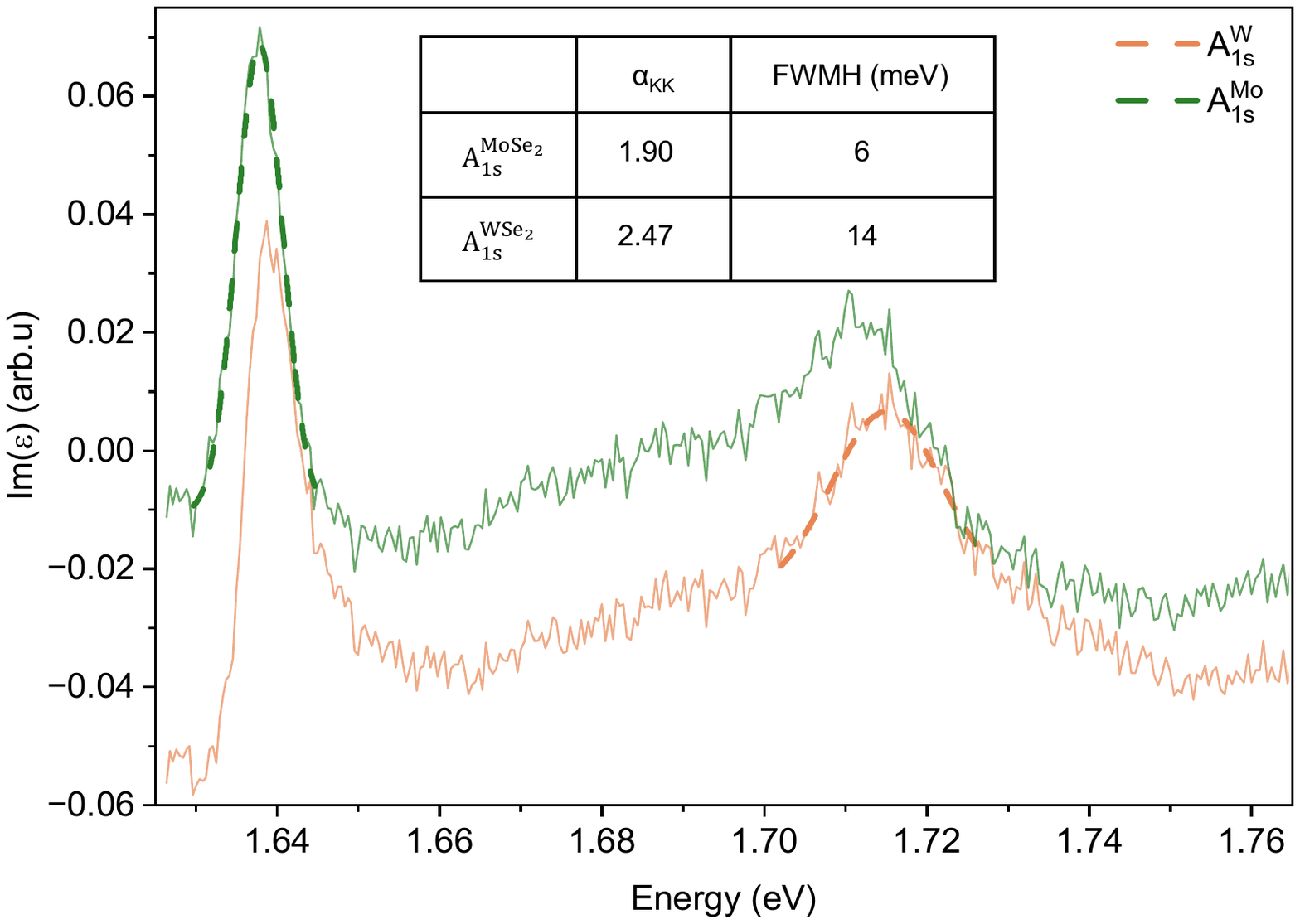}
\caption{Kramers-Kronig transfromed DR/R with Gaussian fits corresponding to the A$_{1s}^{\mathrm{MoSe_2}}$ (green) and A$_{1s}^{\mathrm{WSe_2}}$ (orange). The inset shows the phase parameters $\alpha_{\mathrm{KK}}$ used for the Kramers-Kronig transforms for the two resonances and the extracted FWHM of the gaussian fits.}
\label{fig:FigureS2} 
\end{figure*}

\newpage
\begin{figure*}
\includegraphics[width=0.8\linewidth]{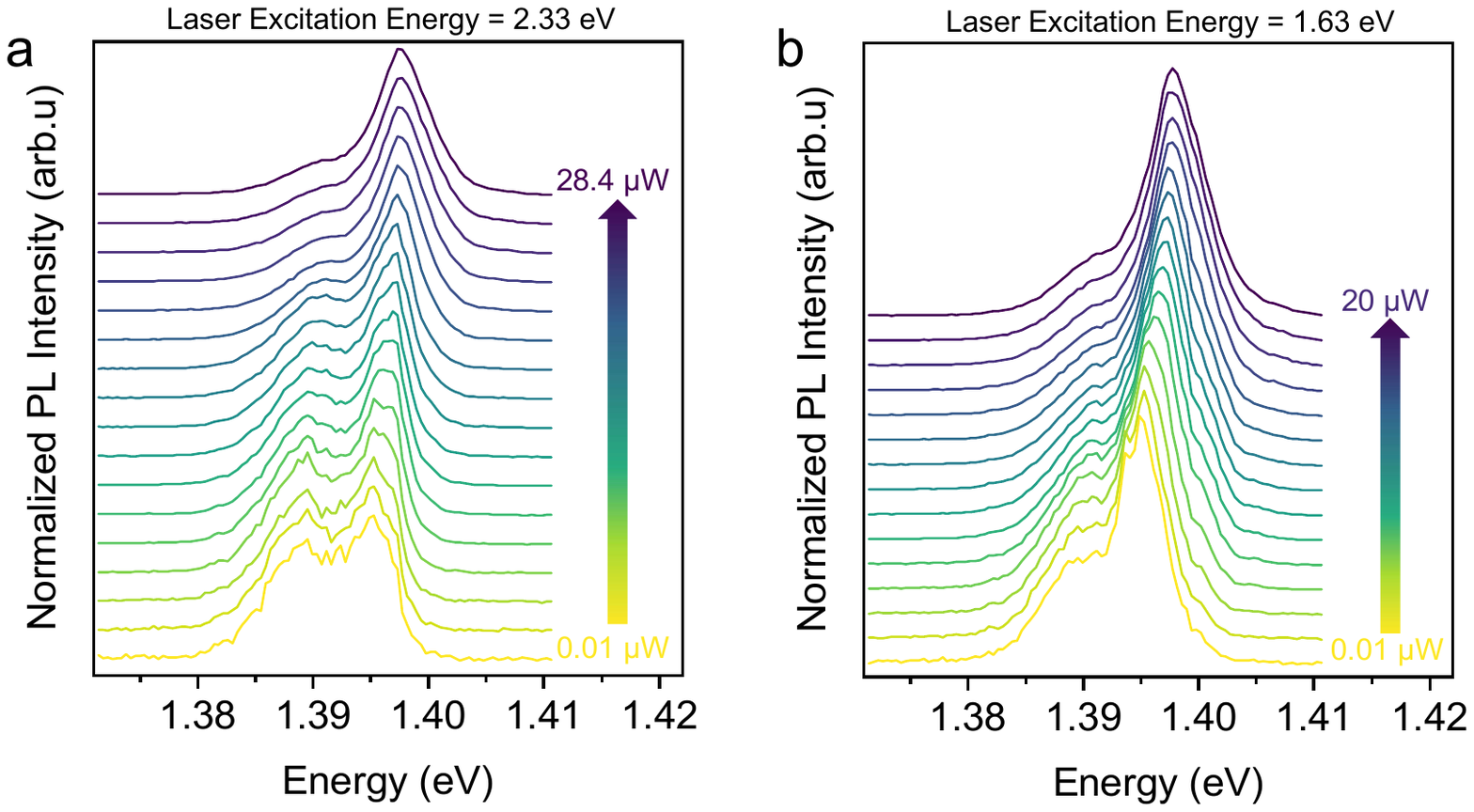}
\caption{Power dependent PL emission from the heterobilayer region with \textbf{a} off-resonant (2.33~eV) and \textbf{b} resonant (1.63~eV) laser excitation energy.}
\label{fig:FigureS3} 
\end{figure*}

\newpage
\begin{figure*}
\includegraphics[width=0.85\linewidth]{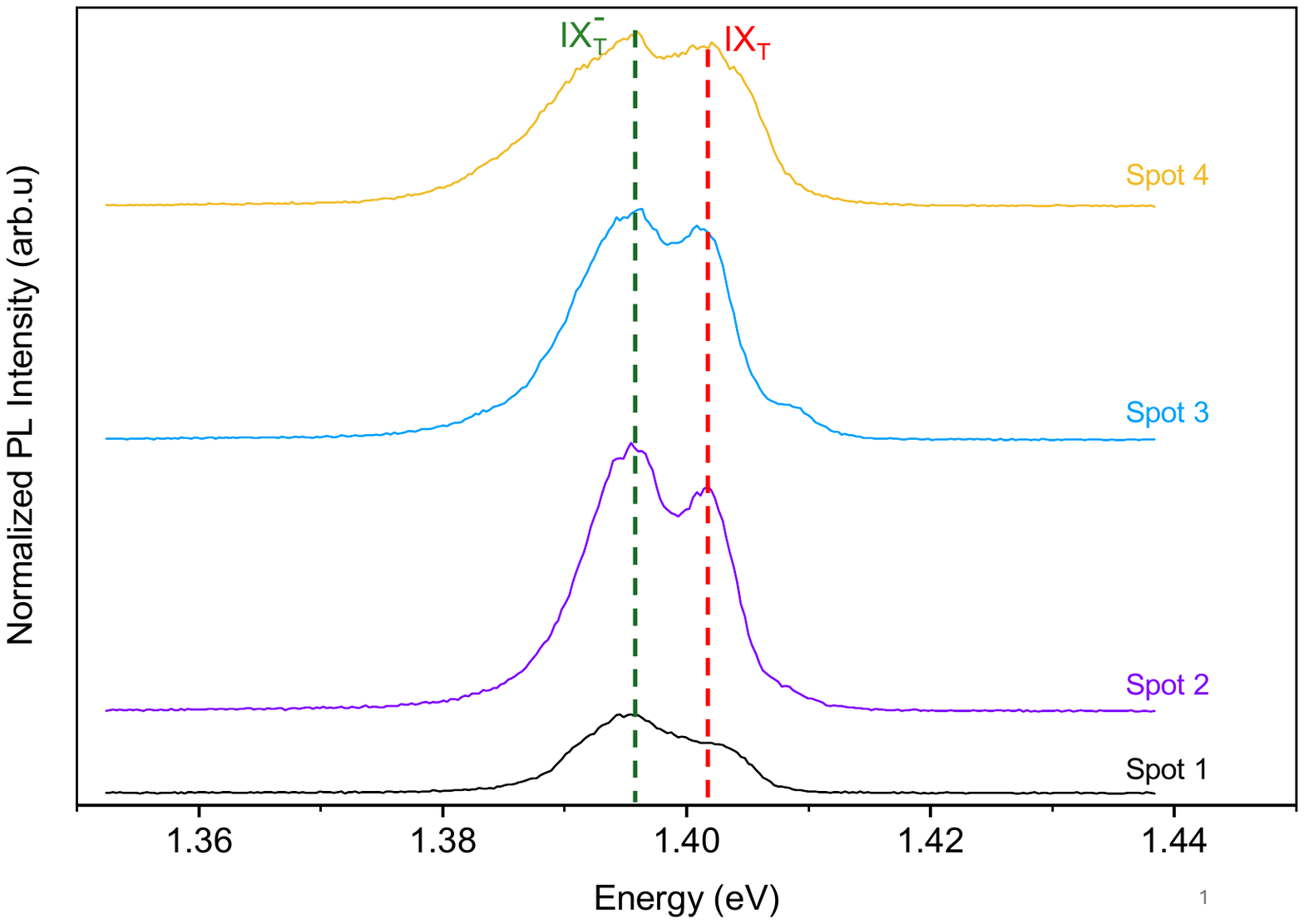}
\caption{Comparison of PL emission at different spots on the heterobilayer region of the encapsulated heterostructure. }
\label{fig:FigureS4} 
\end{figure*}

\end{document}